\documentstyle[12pt]{article}
\begin{document}
\setlength{\unitlength}{1mm}
\renewcommand\theequation{\thesection.\arabic{equation}}
\newcommand{\gapproxeq}{\lower .7ex\hbox{$\;\stackrel{\textstyle >}{\sim}\;$}}
\newcommand{\lapproxeq}{\lower .7ex\hbox{$\;\stackrel{\textstyle <}{\sim}\;$}}
\newcommand{\eqn}[1]{(\ref{#1})}
\newcommand{\be}{\begin{equation}}
\newcommand{\ee}{\end{equation}}
\newcommand{\tnabla}{{\nabla}}
\newcommand{\bea}{\begin{eqnarray}}
\newcommand{\eea}{\end{eqnarray}}
\newcommand{\bean}{\begin{eqnarray*}}
\newcommand{\eean}{\end{eqnarray*}}
\renewcommand{\thefootnote}{\fnsymbol{footnote}}
\newcommand{\del}{\partial}
\newcommand{\I}{\mbox{\rm I} \hspace{-0.5em} \mbox{\rm I}\,}
\newcommand{\R}{\mbox{I \hspace{-0.82em} R}}
\newcommand{\quater}{\mbox{I \hspace{-0.86em} H}}
\def\cstok#1{\leavevmode\thinspace\hbox{\vrule\vtop{\vbox{\hrule\kern1pt
\hbox{\vphantom{\tt/}\thinspace{\tt#1}\thinspace}}
\kern1pt\hrule}\vrule}\thinspace}
\newcommand{\complex}{
	\mbox{C \hspace{-1.16em} \raisebox{-0.018em}{\sf l}}\;}
\newcommand{\ot}{\otimes} 
\newcommand{\otn}{\otimes\I_N} 
\newcommand{\alg}{\mbox{${\cal A}$}}

{\hfill Universit\`a di Napoli Preprint DSF-T-9/96}

\hfill hep-th/9603095\\
\begin{center}
{\Large\bf Constraints on Unified Gauge Theories from 
Noncommutative Geometry}\\
\end{center}

\bigskip\bigskip

\begin{center}
{{\bf F. Lizzi}, {\bf G. Mangano}, {\bf G. Miele} and 
{\bf G. Sparano}}
\end{center}

\bigskip

\noindent
{\it Dipartimento di Scienze Fisiche, Universit\`a di Napoli 
''Federico II'', and INFN, Sezione di Napoli, Mostra D'Oltremare Pad. 20, 
I-80125 Napoli, Italy.}\\

\bigskip\bigskip\bigskip

\begin{abstract}
The Connes and Lott reformulation of the strong and electroweak model
represents a promising application of noncommutative geometry. In this scheme
the Higgs field naturally appears in the theory as a particular {\it gauge
boson}, connected to the discrete internal space, and its quartic potential,
fixed by the model, is not vanishing only when more than one fermion
generation is present. Moreover, the exact hypercharge assignments and
relations among the masses of particles have been obtained. This paper
analyzes the possibility of extensions of this model to larger unified gauge
groups. Noncommutative geometry imposes very stringent constraints on the
possible theories, and remarkably, the analysis seems to suggest that no
larger gauge groups are compatible with the noncommutative structure, unless
one enlarges the fermionic degrees of freedom, namely the number of particles.
\end{abstract}

\vspace{6.cm}

{\footnotesize 
\begin{itemize}
\item[] E-mail:\\
	Lizzi@axpna1.na.infn.it\\
	Mangano@axpna1.na.infn.it\\
	Miele@axpna1.na.infn.it\\
	Sparano@axpna1.na.infn.it
\end{itemize}}
\newpage
\baselineskip=.8cm
\section{Introduction}

There seems to be little doubt that Yang--Mills theories provide the
correct framework to describe physical interactions at the elementary
particle level. Nevertheless the realistic model of the fundamental
interactions (gravity excluded) which has been built according to
these ideas, the so-called standard model (SM), still contains
features which are not completely satisfactory. This would suggest the
presence of  deeper unifications at scales higher than the electroweak
one, based on larger gauge groups. Actually, there are cosmological
open problems, i.e. the baryon asymmetry of the universe, the
inflationary phase and the dark matter problem, which would probably
receive an adequate solution within an extended gauge model. 

Among the unappealing features of the SM there is certainly the large
number of free parameters which should be fixed by experiment, the
necessity of several irreducible representations (IRR's) to describe
all fermions, and the {\it ad hoc} introduction of the Higgs field
with its quartic potential in order to drive the spontaneous symmetry
breaking $SU(3)_C \otimes SU(2)_L \otimes U(1)_Y \rightarrow SU(3)_C
\otimes U(1)_Q$. Moreover, there is an unexplained triplication of
fermion families. The first two of these problems are partially solved
by embedding the SM into a unified model corresponding to a larger
gauge group, such as $SU(5)$, $SO(10)$ or $E_6$. However, in doing so,
there is an increase in the arbitrariness of the model in two
respects: the choice of the unified group, and of a suitable set of
Higgs fields to perform all necessary symmetry breaking. 

Connes and Lott  (CL) have reformulated the SM
\cite{ConnesLott,Connes} using the tools of Noncommutative Geometry
(NCG) \cite{BOOK}, a novel branch of mathematics. Remarkably, in this
scheme the Higgs field naturally emerges as a Yang--Mills field, on
the same footing of the gauge vector bosons, and the quartic
potential, together with the kinetic term, is nothing but its
Yang--Mills action \cite{Varilly}-\cite{KastIochSchuck}. Furthermore,
one gets the correct hypercharge assignments \cite{cordelia}, the
indication that the number of fermion families must be larger than
one, and interesting fuzzy relations among particle masses
\cite{KastIochSchuck}. The vector behaviour of the strong interactions
also emerges naturally \cite{Becca}. NCG models may also provide for
an inflationary phase in the early universe \cite{Napolinfla}. 

On the basis of such an interesting result, it is worth investigating
the unification programme in the framework of NCG. We have done this
analysis under the assumption that the Hilbert space on which the
model is constructed is the space of physical fermionic degrees of
freedom, namely the observed particles. We did however allow for the
presence of right-handed neutrinos. There have been previous attempts
in this direction. Notably Chamseddine, Felder and Fr\"olich
\cite{Zurich1,Zurich2}, have succeeded in building unified theories
akin to the Connes--Lott model. Unlike CL (and the present paper)
however they use an auxiliary Hilbert space which is not the space of
physical fermionic degrees of freedom. 

We have considered both simple and semisimple groups containing the SM
and then applied the CL formalism. In particular, this implies that
the fundamental structure to consider is the smaller algebra of
matrices containing the chosen group as the set of unitary elements,
up to a $U(1)$ factor removed by the unimodularity condition
\cite{ConnesLott,Connes}. The IRR's for these algebras, considered as
real algebras, are only the fundamental one and its complex conjugate.
This rules out all simple groups, like $SU(5)$ or $SO(10)$, for which
fermions belong, in general, to non fundamental IRR's. As a further
constraint on viable models, in the CL approach the Poincar\'e duality
condition \cite{ConnesLott,Connes} has to be satisfied in order to
have gauge invariance. The requirement that both the mentioned
conditions are satisfied leads to the conclusion that there are no
possible extensions even to larger semisimple groups. 

The paper is organized as follows: in section 2 we briefly review the
standard model \'a la Connes--Lott, in the new version of
\cite{Connes}; in section 3 we study all possible extensions of the SM
in the framework of NCG. Finally we give our conclusions in section 4.

\section{The standard model \'a la Connes--Lott \label{standard}}

We will present here a very brief introduction of the {\it new 
version} \cite{Connes} of the CL model. In the following analysis the 
general framework introduced in Refs. \cite{KastIochSchuck} and 
\cite{cordelia} will be adopted. 

In the usual construction of a gauge theory, several ingredients are
required: a space--time manifold $M$, a gauge symmetry group $G$, a
set of fields defined on $M$ and belonging to some IRR's of $G$, and a
lagrangian density, invariant under the gauge group, ruling the
behaviour of such fields on $M$. The fields of the model are divided
into {\it matter fields} (fermionic degrees of freedom), {\it gauge
bosons}, and, where spontaneous symmetry breaking is occurring, {\it
Higgs fields} with their quartic potential. 

The above construction has a geometric interpretation which allows for
a straightforward NCG generalization. In this scheme, the
gauge fields define a connection $1$-form, the ordinary derivatives
are replaced by the covariant ones, and the kinetic term for gauge
bosons is given by the square of the curvature associated to the gauge
connection. 

The programme of NCG \cite{BOOK} is based on the observation that it
is possible to study the properties of a manifold $M$, usually seen as
a geometrical set of points, looking at the algebra of complex
continuous functions defined on it. This opens the perspective for a
noncommutative generalization by considering a noncommutative algebra
${\cal A}$. By observing that the topological, metrical, and
differential properties of the usual pseudo-riemaniann manifolds are
very well captured by the Dirac operator, the basic ingredients of a
noncommutative gauge model can be summarized in terms of the real
spectral triple (${\cal A},{\cal H}, D$). ${\cal A}$ stands for a
$*$-algebra represented on the Hilbert space ${\cal H}$, and $D$
(Dirac operator) is an unbounded selfadjoint operator with compact
resolvent, such that the commutator of $D$ with any element of ${\cal
A}$ is a bounded operator. In the case of the standard model the
algebra ${\cal A}$ is the tensor product of two algebras, ${\cal A} =
C^\infty(M, \complex)\otimes {\cal A}_F$ where $C^\infty(M, \complex)$
is the  algebra of smooth functions on $M$, and ${\cal A}_F =
\quater\oplus\complex\oplus M_3(\complex)$ is the smallest algebra
containing the group to be gauged $G= SU(2) \otimes U(1) \otimes U(3)$
as the set of its unitary elements\footnote{Notice that this group
contains, in addition to the SM one, an extra $U(1)$ factor which can
be removed by applying the unimodularity condition
\cite{ConnesLott,Connes}}. With $\quater$ we denote the algebra of
quaternions, represented as $2 \times 2$ matrices 
\be
\left( \begin{array}{cc} x & -y^* \\ y & x^* \end{array} 
\right) \in \quater~~~~\mbox{with}~~~~x,y\in \complex~~~,
\label{2.1}
\ee
and $M_3(\complex)$ is the algebra of complex $3 \times 3$ matrices. 
The Hilbert space, ${\cal H}$, is the tensor product $ {\cal H} = 
L^2(S_M)\otimes {\cal H}_F$, and is the space of spinor fields 
containing both particles and antiparticles. ${\cal H}_F$  can  
be decomposed according to chirality as     
\be
{\cal H}_F = {\cal H}_L\oplus {\cal H}_R\oplus {\cal H}^c_R\oplus 
{\cal H}^c_L~~~, 
\label{2.2}
\ee
where
\bea
{\cal H}_L & = & 
\left(\complex^2 \otimes \complex^N \otimes \complex^3 \right)
\oplus \left(\complex^2 \otimes \complex^N \otimes \complex \right) ~~~,
\label{2.3}\\
{\cal H}_R & = & 
\left(\left(\complex \oplus \complex\right) \otimes \complex^N 
\otimes \complex^3 \right)
\oplus \left(\complex \otimes \complex^N \otimes \complex \right) ~~~,
\label{2.4}
\eea
and ${\cal H}^c_{L,R}$ are the corresponding spaces for antiparticles
\bea
{\cal H}^c_R & = & 
\left(\complex^2 \otimes \complex^N \otimes \complex^3 \right)
\oplus \left(\complex^2 \otimes \complex^N \otimes \complex \right) ~~~,
\label{2.3a}\\
{\cal H}^c_L & = & 
\left(\left(\complex \oplus \complex\right) \otimes \complex^N 
\otimes \complex^3 \right)
\oplus \left(\complex \otimes \complex^N \otimes \complex \right) ~~~,
\label{2.4a}
\eea

The Hilbert space ${\cal H}_F$ has dimensions equal to the number of
different fermions and antifermions, i.e. $30N$ where $N$ is the
number of generations. For the moment we do not consider right-handed
neutrinos for the SM. A generalization including them and a discussion
of massive Dirac neutrinos can be found in \cite{Pepenu}. Thus, the
natural basis is given by the flavour eigenstate degrees of freedom,
namely, for $N=3$ 
\bea & 
\left(\begin{array}{c} u_{\alpha} \\ d_{\alpha} \end{array}\right)_L~,~
\left(\begin{array}{c} c_{\alpha} \\ s_{\alpha} \end{array}\right)_L~,~
\left(\begin{array}{c} t_{\alpha} \\ b_{\alpha} \end{array}\right)_L~,~
\left(\begin{array}{c} \nu_{e} \\ e \end{array}\right)_L~,~
\left(\begin{array}{c} \nu_{\mu} \\ \mu \end{array}\right)_L~,~
\left(\begin{array}{c} \nu_{\tau} \\ \tau \end{array}\right)_L~,~
\label{2.4b}\\ 
& & \nonumber\\
& \begin{array}{c} (u_{\alpha})_R, \\ (d_{\alpha})_R, \end{array}~
\begin{array}{c} (c_{\alpha})_R, \\ (s_{\alpha})_R, \end{array}~
\begin{array}{c} (t_{\alpha})_R \\  (b_{\alpha})_R, \end{array}~
(e)_R,~(\mu)_R,~(\tau)_R,
\label{2.4c}\\ 
& & \nonumber\\
& \left(\begin{array}{c} u^c_{\alpha} \\ d^c_{\alpha} \end{array}\right)_R~,~
\left(\begin{array}{c} c^c_{\alpha} \\ s^c_{\alpha} \end{array}\right)_R~,~
\left(\begin{array}{c} t^c_{\alpha} \\ b^c_{\alpha} \end{array}\right)_R~,~
\left(\begin{array}{c} \nu_{e}^c \\ e^c \end{array}\right)_R~,~
\left(\begin{array}{c} \nu_{\mu}^c \\ \mu^c \end{array}\right)_R~,~
\left(\begin{array}{c} \nu_{\tau}^c \\ \tau^c \end{array}\right)_R~,~
\label{2.4d}\\ 
& & \nonumber\\
& \begin{array}{c} (u^c_{\alpha})_L, \\ (d^c_{\alpha})_L, \end{array}~
\begin{array}{c} (c^c_{\alpha})_L, \\ (s^c_{\alpha})_L, \end{array}~
\begin{array}{c} (t^c_{\alpha})_L  \\ (b^c_{\alpha})_L, \end{array}~
(e^c)_L,~(\mu^c)_L,~(\tau^c)_L,
\label{2.4e}
\eea
where $\alpha=1,2,3$ is the colour index\footnote{Note that according
to our notations the right-handed antiparticles (\ref{2.4d}) transform
as $\overline{\mbox{{\bf 2}}}$ under $SU(2)_L$, while usually the
doublets obtained from (\ref{2.4d}) by applying $i \sigma_2$ are
considered, transforming as {\bf 2}.}. 

A key point in the CL construction is to introduce a real structure on
the spectral triple (${\cal A},{\cal H},D$). We first consider the two
linear isometries $\epsilon$, which has the spaces of particles and
antiparticles as eigenspaces with eigenvalues respectively equal to
$1$ and $-1$, and the chirality $\chi$. These two operators realize
the decomposition of the Hilbert space as in Eq. \eqn{2.2} and yield
the following properties 
\be                                                          
\epsilon^2 = 1~,~~~ \chi^2 = 1~,~~~ \chi = \chi^{\dagger}~,\label{2.4f}
\nonumber
\ee
\be
\left[\epsilon,\chi\right] = 0~,~~~ [\alpha, \chi] =
[\alpha,\epsilon] = 0~,~\forall 
\alpha \in {\cal A}~,~~~
\left\{ \chi, D \right\} = 0~,\label{2.6}
\ee
A real structure is then an antilinear isometry $J$ of ${\cal H}$ 
satisfying the following properties
\be                                                         
J^2 = 1~,~~~ \{ \epsilon, J \} = [J,D] = 0~,~~~ J \chi = \pm \chi J~~~, 
\label{2.7}
\ee
\be
\left[ \alpha, J \beta J^{\dag}\right] =0~,~~~ 
\left[\left[D,\alpha\right],J \beta J^{\dag}\right] = 0~, ~~~
 \forall \alpha,\beta \in {\cal A}~~~,  \label{2.8}
\ee
where $\{~,~\}$ denotes anticommutation\footnote{The introduction of
such an operator is inspired by Tomita's theorem \cite{tomita,take},
which, for a Von Neumann algebra with cyclic and separating vector in
${\cal H}$, gives an antilinear involution such that $ J {\cal A}
J^{\dagger} $ is the commutant of the algebra.}. The last two
equations (\ref{2.8}), as explained in \cite{Connes}, are related to
Poincar\'e duality and ensure the gauge invariance of the lagrangian
density. 

In the following we will restrict our analysis to the finite part of
the triple (${\cal A}_F, {\cal H}_F, D_F$), where $D_F$ is defined
from the Dirac operator $D$ by the relation \cite{cordelia} 
\be
D=\partial\!\!\!/ \ \otimes \I + \I \otimes D_F~~~.\label{2.8a}
\ee
 
Let us now consider an element $(a,b,c)$ of ${\cal A}_F$, where $a$,
$b$ and $c$ belong to $\quater$, $\complex$ and $M_3(\complex)$
respectively. A faithful representation $\rho$ of ${\cal A}_F$ on the
Hilbert space ${\cal H}_F$ is the following 
\be
\rho(a,b,c) \equiv \left(\begin{array}{cc}\rho_{w}(a,b) & 0\\
0 & \rho^*_s(b,c) \end{array} \right)~~~,\label{2.9}
\ee
where
\bea
\rho_{w}(a,b) & \equiv & \left( \begin{array} {cccc}
a \otimes \I_N \otimes \I_3 & 0 & 0 & 0 \\
0 & a \otimes \I_N & 0 & 0\\
0 & 0 & B \otimes \I_N \otimes \I_3 & 0 \\
0 & 0 & 0 & b^*~ \I_N \end{array} \right),~~~B \equiv \left( 
\begin{array}{cc}b & 0\\
0 & b^* \end{array}\right) \label{2.10}\\
\nonumber\\ 
\rho_{s}(b,c) & \equiv & \left( \begin{array} {cccc}
\I_2 \otimes \I_N \otimes c & 0 & 0 & 0 \\
0 & b^*~ \I_2 \otimes \I_N & 0 & 0\\
0 & 0 & \I_2 \otimes \I_N \otimes c & 0 \\
0 & 0 & 0 & b^* ~\I_N \end{array} \right). \label{2.10a} 
\eea

The operators $\chi_F$ and $J_F$ have the form
\bea
\chi_F &=& \left( \begin{array} {cccc}
-\I_{24} & 0 & 0 & 0 \\
0 & \I_{21} & 0 & 0\\
0 & 0 & \I_{24} & 0 \\
0 & 0 & 0 & - \I_{21} \end{array} \right), \label{2.11}  \\
\nonumber\\ 
J_F & = & J^{\dagger}_F = 
\left( \begin{array}{cc} 0 & \I_{45}\\
\I_{45} & 0 \end{array}\right)~C~~,\label{2.12}\\
\eea
where $C$ is the complex conjugation. The role of $J_F$ is to
interchange particles with antiparticles and, at same time, chirality.
It therefore acts, up to a complex conjugation, as the Dirac charge
conjugation ${\cal C}$ 
\be
{\cal C} \psi_{L,R} = i \gamma_2 \psi_{L,R}^* = (\psi^c)_{R,L}~~~,
\ee
where $\gamma_2$ is the second Dirac matrix.

The euclidean Dirac operator $D_F$ for the standard 
model is
\begin{equation}
D_{F} = \left( \begin{array} {cccc}
0 & {\cal M} & 0 & 0 \\
{\cal M}^{\dag} & 0 & 0 & 0\\
0 & 0 & 0 & {\cal M^*} \\
0 & 0 & 
{\cal M}^{T} & 0 \end{array} \right). 
\label{2.13}  \\
\end{equation}
The $24 \times 21$ mass matrix ${\cal M}$ is given by
\be
{\cal M} =  \left( \begin{array} {cc} \left( \begin{array} {cc} 
	  M_u\otimes\I_{3} &  0  \\
			0 & M_d\otimes\I_{3}   \end{array} \right) & 0  \\ 
		 0 & \left( \begin{array} {c} 0 \\ M_e  
\end{array} \right) \end{array} \right) , \label{2.14}  
\ee
where
\bea
M_u = \left( \begin{array}{ccc} 
m_u& 0 & 0 \\
0 & m_c & 0 \\
0 & 0 & m_t\\
\end{array} \right),~~
M_d = C_{KM} \left( \begin{array}{ccc} 
m_d& 0 & 0 \\
0 & m_s & 0 \\
0 & 0 & m_b\\
\end{array} \right),~~
M_e = \left( \begin{array}{ccc} 
m_e& 0 & 0 \\
0 & m_\mu & 0 \\
0 & 0 & m_\tau\\
\end{array} \right). \nonumber\\ \label{2.15}
\eea
In the previous relations $C_{KM}$ denotes the
Cabibbo--Kobayashi--Maskawa mixing matrix. Notice that the
antiparticle mass matrix appearing in the lower right corner of
(\ref{2.13}) is obtained from the corresponding one for particles by
using the charge conjugation on bilinear mass terms in the lagrangian
density 
\be
\left( \begin{array} {cc}
0 & {\cal M} \\
{\cal M}^{\dag} & 0\\ \end{array} \right) 
\stackrel{{\cal C}}{\rightarrow} 
\left( \begin{array} {cc}
0 & {\cal M^*} \\
{\cal M}^{T} & 0\\ \end{array} \right) ~~~.
\ee

In general, given a Dirac operator $D$, the gauge connection is written as
\begin{equation}
A = \sum_{i} \beta_i [D,\alpha_i] \equiv \sum_{i} \beta_i d \alpha_i~~~,
\label{2.16}
\end{equation}
where $\alpha_i$ and $\beta_i$ are elements of ${\cal A}$, and the
differential $d$ is defined by $d \alpha \equiv [D,\alpha]$. From the
connection $A$, one defines the curvature $\theta$ as\footnote{Note
that the $d$ operator so defined is not nilpotent and hence a
quotient is necessary in order to to obtain the correct differential
algebras \cite{BOOK,Varilly}.} 
\begin{equation}
\theta \equiv dA + A^2~~~,\label{2.17}
\end{equation}
and thus the bosonic lagrangian density is obtained as 
\begin{equation}
{\cal L_{B}} = { 1 \over {\cal N}} \mbox{Tr}\,\theta^2~~~,\label{2.18}
\end{equation}
where ${\cal N}$ is a normalization constant.

In order to define the fermionic part of the Lagrangian, let us observe that 
the operator $J$ enables the definition of a right action of the
algebra ${\cal A}$ on ${\cal H}$ as
\be
\Psi\alpha\equiv J\alpha^{\dag} J\Psi~~~.\label{2.19}
\ee
The first relation in (\ref{2.8}) ensures that right and left actions
commute. We can now define an adjoint action of the gauge group,
identified with the set of unitary elements of the algebra, on the
Hilbert space 
\be
{}^u\Psi=u\Psi u^{\dag}=uJuJ\Psi~~~,\label{2.20}
\ee
and the fermionic lagrangian density is
\be
\Psi^{\dag}(D+A+JAJ)\Psi~~~.\label{2.22}
\ee
Using relations (\ref{2.8}) it is possible to check \cite{cordelia} that
(\ref{2.22}) is invariant under the transformations (\ref{2.20}) and
\be
{}^uA=uAu^{\dag}+u[D,u^{\dag}]~~~.\label{2.23} 
\ee
For the Dirac operator $D$ defined in Eqs. (\ref{2.8a}), (\ref{2.13})
and the algebra $C^{\infty}(M,\complex)\otimes {\cal A}_F$, where
${\cal A}_F$ is represented as in (\ref{2.9})--(\ref{2.10a}), one
obtains the full Lagrangian of the SM. In particular, the $SU(2)_L$
doublet Higgs field, $\varphi$, naturally arises along with its
kinetic term and quartic potential 
\be
V(\varphi)={K\over 16 L^2}|\varphi|^4 - {K\over 2L}|\varphi|^2~~~,
\label{2.24}
\ee
where $K$ and $L$ are known functions of the fermion masses
\cite{KastIochSchuck}. From the lagrangian density one can obtain some
relations among particle masses and coupling constants of SM
\cite{KastIochSchuck}. In particular, by denoting with $g_2$ and $g_3$
the gauge couplings for $SU(2)_L$ and $SU(3)_C$ respectively, one has 
\begin{eqnarray}
\sin^2 \theta & < & {2 \over 3} \left( 1 + { 1 \over 9} \left(
{g_2 \over g_3 } \right)^2 \right)^{-1}~~~,\label{2.25}\\
m_e^2 & < & m_W^2 < {1 \over 3} \left( m_t^2 + m_b^2 + m_c^2 + m_s^2 + m_d^2
+ m_u^2\right) \approx  {1 \over 3} m_t^2~~~,\label{2.26}\\
m_{H} & = & (280 \pm 33)~GeV~~~.\label{2.27}
\end{eqnarray}
Note that (\ref{2.26}) provides a lower bound for the {\it top} quark
mass, and (\ref{2.27}) has been obtained from a relation which
involves the quark masses ({\it top} included). 

\section{Unified theories in noncommutative geometry}

We have seen that in the CL approach it is possible to obtain a
complete description of the SM and, in particular, of the spontaneous
symmetry breaking mechanism down to $SU(3)_C\ot U(1)_Q$ by suitably
choosing the structure of the Dirac operator. In this section we will
analyze if this approach is compatible with larger gauge group
symmetries, which are effective at higher energies, and which at some
scale break down to the SM. It is well known that much effort has been
devoted to the so called unification programme, namely to find simple
gauge groups which contains the SM and are compatible with the low
energy phenomenology. The simplest version of the non supersymmetric
Georgi and Glashow model, based on the group $SU(5)$, has been ruled
out by the accurate experimental results on the strong coupling
constant, $\sin^2 \theta_W$ at the $M_Z$ scale, and the lower limit on
the proton lifetime. Among the groups whose algebra have rank 5,
unification based on $SO(10)$ is still consistent with all
experimental constraints and also gives interesting predictions for
neutrino masses. This latter point is particularly relevant for many
topics which are at the border between particle physics and cosmology,
like the solar neutrino problem and the nature of dark matter. Many
other attempts have been done by considering, for example, exceptional
algebras, or larger unitary and orthogonal groups, in which the
generation degrees of freedom are gauged\footnote{For a review see
\cite{unif}.}. It is also worth mentioning that in this unification
programme, several semisimple groups have been studied as well, which
would represent an intermediate step towards a complete unification.
Typical examples are the left-right models, as $SU(4)_{PS} \ot SU(2)_L
\ot SU(2)_R$ or $SU(3)_C \ot SU(2)_L \ot SU(2)_R \ot U(1)_{B-L}$,
where $SU(4)_{PS}$ is the Pati-Salam group \cite{Pati}, which for
example appears as possible intermediate symmetry stages in the
$SO(10)$ breaking to the SM. 

In this section we will consider all possible simple and semisimple
algebras, which contain the SM, and for which the CL geometrical point
of view will be applied. It is remarkable that this approach gives
more constraints than the ones which should be satisfied if one adopts
the customary way to construct a spontaneously broken gauge theory.
Usually the IRR's which are used to represent fermionic fields are, in
general, not constrained to be the fundamental ones of the considered
Lie algebras. However, as we have seen, in the CL approach one starts
with the smallest algebra of matrices which contains the chosen gauge
group as the set of unitary elements. The only IRR's of the gauge
group which are allowed are the ones coming from corresponding IRR's
of the algebra. It follows that {\it only the fundamental IRR and its
complex conjugate can be used to classify fermions}. 

In the analysis we will impose several requirements, some of which are
quite obvious and usually assumed. Others, as the one on the
dimensionality of IRR's just discussed, are instead more related to
the CL construction. In particular: 
\begin{itemize}
\item[{\it i})] the algebra should contain $\quater \oplus \complex \oplus 
M_3(\complex)$;
\item[{\it ii})] we will consider only the usual particle spectrum, the one 
already present in the SM, with the only possible
addition of right-handed neutrinos;
\item[{\it iii})] only {\em fundamental} IRR's can be 
used to accomodate left and right-handed fermions; 
\item[{\it iv})] the IRR's should be {\em complex} in order to allow
left and right-handed particles to behave independently under the 
action of the gauge group;
\item[{\it v})] the IRR's should contain only colour triplets 
and weak isospin doublets 
of $SU(2)_L$;
\item[{\it vi})] all components of the gauge connection, namely gauge vector
bosons {\it and}  Higgs fields, should transform as
\begin{equation}
A \rightarrow  u A u^{\dag} + u[D,u^{\dag}]~~~.
\label{contra}
\end{equation}
\end{itemize}
Notice that, commonly there are no constraints on the IRR's to be used
for Higgs fields but general symmetry requirements on the lagrangian
density. The transformation rule (\ref{contra}) for the Higgs is
peculiar of their geometrical interpretation as a part of the
connection on the noncommutative components of the geometrical space.
We will stress in the following that (\ref{contra}) is consistent with
gauge invariance of the interaction terms among fermions and gauge
bosons only if the conditions (\ref{2.8}) are satisfied. 

Condition {\it iv}) can be motivated as follows. In discussing unified
theories it is convenient to use as a basis for matter fields fermions
and antifermions both left-handed, belonging to one or more IRR's of
the gauge group. If we denote this basis with $f_L$, the right-handed
fermions $f_R$ would transform then as $f_L^*$. Thus, if we had chosen
a self-conjugate IRR of the gauge group right and left-handed
particles would had transformed in the same way. 

Conditions {\it iii}) and {\it iv}) rule out the possibility to use
orthogonal groups for unified models, since $SO(n)$ fundamental IRR's
are all self-conjugate. In particular, it is worth noticing that
unification based on $SO(10)$, which is the most appealing non
supersymmetric unified model, cannot be realized in the CL approach.
For the above reasons we will not consider in the following analysis
orthogonal groups. 

We point out that all our discussion is still at the classical level.
It is well known that appearance of anomalies spoils gauge invariance
at quantum level\footnote{The cancellation of anomalies for the SM has
an interesting counterpart in the unimodularity condition in CL
\cite{Madridanom}.}. This imposes additional constraints on viable
unified gauge models. However from our analysis, it appears that
already there are no possible choices satisfying all {\it classical}
requirements $i)-vi)$. 

We will not consider supersymmetric theories. It would be interesting
to extend a similar analysis also to this case. We will start by
discussing simple groups and then we will consider the semisimple
ones, ordered with increasing rank. In our notations, an IRR for the
group $G_1 \ot ... \ot G_n$ is denoted by $({\mbox{\bf d}_1},
...,{\mbox{\bf d}}_n)$, where ${\mbox{\bf d}}_i$ is the
$d_i$--dimensional IRR of $G_i$. 
 
\subsection{Simple Groups}

\noindent
{\it Rank 4: $SU(5)$}\\
The left-handed fermions are accommodated in two IRR's;
$\overline{\mbox{\bf 5}} \oplus \mbox{\bf 10}$. The $\mbox{\bf 10}$
can be obtained as the antisymmetric part in the product $\mbox{\bf 5}
\otimes \mbox{\bf 5}$. Choosing the algebra as $M_5(\complex)$, the
$\mbox{\bf 10}$ would not be a IRR of the algebra but only of its
group of unitaries. This is a typical case in which Yang--Mills
theories have a freedom that Connes--Lott models do not have. 
 
\noindent
{\it Rank 14 and 15: $SU(15)$ and $SU(16)$}\\
Let us first consider $SU(16)$. For the algebra $M_{16}(\complex)$
left--handed fermions can form a $\mbox{\bf 16}$, while the
right--handed ones a $\overline{\mbox{\bf  16}}$, which are complex
representations as they should. The problem arises however for the
embedding 
\be
SU(3)_C \otimes SU(13) \otimes U(1) \subset SU(16) \label{subsu16}~~~.
\ee
In this case the $\mbox{\bf 16}$ decomposes as
\be
\mbox{\bf 16} = (\mbox{\bf 3},\mbox{\bf 1}, {\bf \lambda})\oplus
(\mbox{\bf 1},\mbox{\bf 13}, -{3\over 13}\lambda)~~~,
\label{16dec}
\ee
and this means that coloured states are $SU(2)_L$ singlets, since
\be
SU(2)_L\subset SU(13)~~~.\label{emb}
\ee
For $SU(15)$ the discussion is analogous. Instead of \eqn{subsu16} we have
\be
SU(3)_C \otimes SU(12) \otimes U(1) \subset SU(15)~~~,\label{emb1}
\ee
and again we find that colour triplets would be $SU(2)_L$ singlets,
which is incorrect. 

\subsection{Semisimple Groups}

\noindent
{\it Rank 5: $SU(4)_{PS}\otimes SU(2)_L \otimes SU(2)_R$}\\
This group was introduced as a relevant example of left--right
symmetric model \cite{Pati}. It also appears as an intermediate stage
in  the symmetry breaking of $SO(10)$ to the SM. In this model, a
fermionic family is accommodated in the following IRR's of the group 
\bea
\left(
\begin{array}{cc}
u_\alpha & \nu_e\\
d_\alpha & e
\end{array}
\right)_L
=
(\mbox{\bf 4},\mbox{\bf 2},\mbox{\bf 1})~~~& , &
\left(
\begin{array}{cc}
{u^c}_\alpha & \nu^c_e\\
{d^c}_\alpha & e^c
\end{array}
\right)_L
=
(\overline{\mbox{\bf 4}},\mbox{\bf 1},\mbox{\bf 2})~~~,
\nonumber\\
\left(
\begin{array}{cc}
u_\alpha & \nu_e\\
d_\alpha & e
\end{array}
\right)_R
=
(\mbox{\bf 4},\mbox{\bf 1},\mbox{\bf 2})~~~&,&
\left(
\begin{array}{cc}
{u^c}_\alpha & \nu^c_e\\
{d^c}_\alpha & e^c
\end{array}
\right)_R
=
(\overline{\mbox{\bf 4}},\mbox{\bf 2},\mbox{\bf 1})~~~.
\label{class422}
\eea
The model, in its minimal version, requires two Higgs multiplets
transforming respectively as $(\mbox{\bf 10},\mbox{\bf 1}, \mbox{\bf
3})$ and $(\mbox{\bf 1},\mbox{\bf 2}, \mbox{\bf 2})$. The first one
breaks the symmetry from $SU(4)_{PS}\otimes SU(2)_L \otimes SU(2)_R$
to SM, whereas the second multiplet drives the breaking to $SU(3)_C
\ot U(1)_Q$. In general the Yukawa coupling of the Higgs multiplet
$(\mbox{\bf 10},\mbox{\bf 1},\mbox{\bf 3})$ to right-handed fermions
provides Majorana mass terms to right-handed neutrinos. 

In order to describe this model in the Connes-Lott framework one
chooses the algebra $M_4(\complex) \oplus\quater_L\oplus\quater_R$.
This is also consistent with the particle content of the model, since
in the classification of fermion families presented in
(\ref{class422}), particles belong to fundamental IRR's of the group
only. On the Hilbert space 
\be
{\cal H}_F=\bigoplus^{N}_{i=1}
\left((\mbox{\bf 4},\mbox{\bf 2},\mbox{\bf 1})\oplus
(\mbox{\bf 4},\mbox{\bf 1},\mbox{\bf 2}) \oplus
(\overline{\mbox{\bf 4}},\mbox{\bf 2},\mbox{\bf 1}) \oplus
(\overline{\mbox{\bf 4}},\mbox{\bf 1},\mbox{\bf 2})\right)_i
\label{4.1}
\ee
(where $i$ stands for the family index), and denoting with $h$, $a_L$
and $a_R$ elements of $M_4(\complex)$, $\quater_L$ and $\quater_R$
respectively, the algebra ${\cal A}_F$ is represented as follows 
\be
\rho(h,a_L,a_R) \equiv \left(
\begin{array}{cccc}
a_L\ot\I_4\otn & & & \\
& a_R \ot \I_4\otn & & \\
& & \I_2 \otimes h^* \otn & \\
& & & \I_2 \otimes h^* \otn 
\end{array}
\right)~~~.\label{4.2}
\ee
In the first phase transition the Yukawa terms provide a Majorana mass
to right-handed neutrinos, while the second phase transition gives
Dirac masses to all particles. It is therefore quite natural to write
the finite component of the Dirac operator $D_F$ as follows 
\begin{equation}
D_{F} = \left( \begin{array} {cccc}
0 & {\cal M} & 0 & 0 \\
{\cal M}^{\dag} & 0 & 0 & \mu\\
0 & 0 & 0 & {\cal M^*} \\
0 & \mu^{\dag} & 
{\cal M}^{T} & 0 \end{array} \right)~~~, 
\label{4.3}  \\
\end{equation}
where 
\be
{\cal M} =  \left( \begin{array} {cc} M_{U} \otimes \I_4 & 0 \\
0 & M_{D} \otimes \I_4 \end{array} \right)~~~. \label{4.4}  
\ee
where $M_{U}$ and $M_{D}$ are the $N \times N$ mass matrices in the
generation space for {\it Up particles} ($u,c,t$ quarks and neutrinos)
and {\it Down particles }($d,s,b$ and charged leptons), respectively.
As far as the Majorana mass matrix $\mu$, it has entries in the
neutrino sector only and can be written as 
\be
\mu =  \left( \begin{array} {ccc} 0_{18} &  & \\ 
 & \mu_{\nu} & \\ & & 0_3 \end{array} \right)~~~, \label{4.5}  
\ee
where with $0_n$ we have indicated a $n \times n $ null matrix, and
$\mu_{\nu}$ is the $3 \times 3$ Majorana mass matrix for neutrinos.
However, the presence of $\mu$, and in general of elements in $D_F$
connecting the particle sector with the antiparticle one, makes
impossible to satisfy the second relation (\ref{2.8}), as it can be
easily checked. This means that the gauge invariance is not guaranteed
\cite{Connes}, and one can actually check by an explicit calculation
that it is violated. This can be easily seen by studying the way the
Higgs bosons corresponding to the choice (\ref{4.3}) transform under a
gauge transformation. Using (\ref{2.16}) the connection in the finite
component of the complete triple results to be 
\begin{equation}
\left( \begin{array} {cccc} 0 & \phi & 0 & 0 \\ 
\phi^{\dagger} & 0 & 0 & \Phi \\
0 & 0 & 0 & 0 \\ 0 & \Phi^{\dagger} & 0 & 0 \end{array} \right)~~~,
\label{discon}
\end{equation}
where we have used the fact that $[M_{U,D} \otimes \I_4 , \I_2 \otimes
h^* \otimes \I_N]=0$ According to (\ref{contra}) one gets the
behaviour of $\phi$ and $\Phi$ under a gauge transformation 
\bea
\phi & \rightarrow & -{\cal M} + a_R^u \, {\cal M} \, 
a_L^{u*} + a_R^u \, \phi \, a_L^{u*} 
\nonumber \\
\Phi & \rightarrow & - \mu + a_R^u \, \mu \, h^u + a_R^u \, \Phi \, h^u
\eea
where $a_L^u$, $a_R^u$ and $h^u$ are unitary elements of $\quater_L$,
$\quater_R$ and $M_4(\complex)$ respectively. From the previous
results, it follows that $\phi$ transforms as a (\mbox{\bf
1},\mbox{\bf 2},\mbox{\bf 2}) IRR under the gauge group, and, as we
have mentioned, it is the Higgs needed to break the symmetry from the
SM down to $SU(3)_C \otimes U(1)_Q$. The multiplet $\Phi$, instead,
transforms as (\mbox{\bf 4},\mbox{\bf 1},\mbox{\bf 2}) and not as
$(\mbox{\bf 10},\mbox{\bf 1},\mbox{\bf 3})$. Only the latter can be
coupled to right-handed fermions and left-handed antifermions in a
gauge invariant way via Yukawa terms, while similar terms involving
$\Phi$ just obtained would spoil gauge invariance of the lagrangian
density. It follows that the breaking of the second of conditions
(\ref{2.8}), due to the introduction of $\mu$, i.e. of Majorana mass
terms, leads to a model which does not satisfy gauge invariance. 

\noindent
{\it Rank 5: $SU(3)_C\otimes SU(2)_L \otimes SU(2)_R \otimes U(1)_{B-L}$}\\
This case is very similar to the previous one. The algebra is
$M_3(\complex)\otimes\quater_L\otimes\quater_R\otimes\complex$, and
the particle assignments in the various representations are as in the
standard model with the only difference that right-handed particles,
including neutrinos, are classified in doublets under $SU(2)_R$ and
singlet under $SU(2)_L$, the reverse being true for left--handed
particles. 

Given $a_L \in \quater_L$, $a_R \in \quater_R$, $c\in M_3(\complex)$
and $b\in\complex$ the representation of the algebra is similar to the
one of the standard model 
\be
\rho(a_L,a_R,b,c) \equiv \left(\begin{array}{cc}\rho_{w}(a_L,a_R) & 0\\
0 & \rho^*_s(b,c) \end{array} \right)~~~,
\ee
where
\bea
\rho_{w}(a_L,a_R) & \equiv & \left( \begin{array} {cccc}
a_L \otimes \I_N \otimes \I_3 & 0 & 0 & 0 \\
0 & a_L \otimes \I_N & 0 & 0\\
0 & 0 & a_R \otimes \I_N \otimes \I_3 & 0 \\
0 & 0 & 0 & a_R\otimes \I_N \end{array} \right),
\nonumber\\ 
\rho_{s}(b,c) & \equiv & \left( \begin{array} {cccc}
\I_2 \otimes \I_N \otimes c & 0 & 0 & 0 \\
0 & b^*~ \I_2 \otimes \I_N & 0 & 0\\
0 & 0 & \I_2 \otimes \I_N \otimes c & 0 \\
0 & 0 & 0 & b^*\I_2\otimes \I_N \end{array} \right).  
\eea
The Dirac operator $D_F$ has the same form as in \eqn{4.3} with $\mu$ given by
Eq. (\ref{4.5}) and 
\be
{\cal M} = \left( \begin{array} {cc} \left( \begin{array} {cc}
M_u \otimes \I_3 & 0 \\ 0 & M_d \otimes \I_3 \end{array} \right) & 0 \\
0 & \left( \begin{array} {cc}
M_{\nu} & 0 \\ 0 & M_e \end{array} \right)
\end{array} \right)~~~.
\ee
Note that $M_u$, $M_d$ and $M_e$ are analogously defined as in
(\ref{2.15}), and $M_{\nu}$ is a neutrino Dirac mass matrix (with
possible mixing). Again the Majorana mass terms spoil the second of
condition (\ref{2.8}) and so also this model is not viable in the
strict CL framework. By reasoning as in the previous case it is easy
to show that the two Higgs multiplets $\phi$ and $\Phi$ of Eq.
(\ref{discon}) transforms under $SU(3)_C \otimes SU(2)_L \otimes
SU(2)_R$ as, respectively, $(\mbox{\bf 1},\mbox{\bf 2},\mbox{\bf 2})$
and $(\mbox{\bf 1},\mbox{\bf 1}, \mbox{\bf 2})$. The $\phi$ multiplet
has the correct behaviour under the gauge group, while the
introduction of Yukawa terms for $\Phi$ explicitly breaks the gauge
invariance of the lagrangian density. We remind the reader that, in
the usual approach, the minimal choice for the Higgs sector of the
$SU(3)_C \otimes SU(2)_L \otimes SU(2)_R \otimes U(1)_{B-L}$ model
involves instead two IRR's transforming as $(\mbox{\bf 1},\mbox{\bf
2}, \mbox{\bf 2})$, which we also got in the CL approach, and
$(\mbox{\bf 1},\mbox{\bf 1},\mbox{\bf 3})$, which instead does not
naturally emerge from it. 

\noindent
{\it Rank 6: $SU(4)\otimes SU(4)$}\\
We start choosing the algebra and the left fermions IRR's as,
respectively, $M_4(\complex) \oplus M_4(\complex)$ and ({\bf 4},{\bf
4}), ($\overline{\mbox{\bf 4}}$,{\bf 4}), or ({\bf
4},$\overline{\mbox{\bf 4}}$). For the embedding 
\be
SU(3)_C\otimes U(1)\subset SU(4) ~~~,
\ee
the 4-dimensional IRR decouples as
\be
\mbox{\bf 4}=(\mbox{\bf 1}, 3 \lambda) + (\mbox{\bf 3},-\lambda)~~~.
\ee
If $SU(3)$ is contained in the first $SU(4)$, then $SU(2)$ must be
contained in the second one. The two possible embedding corresponding
to maximal subalgebras are 
\begin{itemize}
\item[$i)$] $SU(2)_L\otimes SU(2)\subset SU(4)$. In this case {\bf 4}=({\bf 
2},{\bf 2}), and this would mean that the fermions would be doublets also under 
the second $SU(2)$ factor.
\item[$ii)$] $SU(2)_L\otimes SU(2)\otimes U(1)\subset SU(4)$. 
In this case {\bf 4}=({\bf 2},{\bf 1},$\lambda$)
+({\bf 1},{\bf 2},$-\lambda$).
\end{itemize}
In both cases only quarks or antiquarks can be accommodated,
respectively, in the $\mbox{\bf 4}$ or $\overline{\mbox{\bf 4}}$
IRR's, since only the {\bf 3} or $\overline{\mbox{\bf 3}}$ of
$SU(3)_C$ is present. The 16 particle states, both left and
right-handed,  would be accommodated into the ({\bf 4},{\bf 4}), while
the antiparticles in the ($\overline{\mbox{\bf 4}},\overline{\mbox{\bf
4}}$). Left and right fermions would therefore appear in the same IRR
and do not transform independently, but would mix under a gauge
transformation. 

\noindent
{\it Rank 8: $SU(8)\otimes SU(2)$}\\
The algebra in this case is $M_8(\complex) \oplus \quater$. The IRR's
can be accordingly chosen as ({\bf 8},{\bf 2}) or
($\overline{\mbox{\bf 8}}$,{\bf 2}), which contain all left or
right-handed fermions, or ({\bf 8},{\bf 1}) and ($\overline{\mbox{\bf
8}}$,{\bf 1}). We cannot choose the $SU(2)_L$ of the SM to be the
$SU(2)$ factor, because in this case fermions would be all doublets or
singlets. Therefore both factors $SU(3)_C$ and $SU(2)_L$ should be
contained in $SU(8)$. For the embedding 
\be 
SU(3)_C \oplus SU(5) \oplus U(1) \subset SU(8)~~~,
\ee
the {\bf 8} decomposes as
\be 
( {\bf 3}, {\bf 1}, 5 \lambda ) \oplus ( {\bf 1} , 5, -3 \lambda)~~~. 
\ee
Since $SU(2)_L \subset SU(5)$ all coloured states would be weak isospin
singlets.

\section{Conclusions}

In this paper we have analyzed the possibility to construct a Grand
Unified Theory in the strict framework of the new version of the
Connes-Lott model. We have assumed the minimal Hilbert space, made of
the degrees of freedom of the fermionic particles already observed,
allowing only for the existence of right-handed neutrinos. Since the
CL model requires the fermions to be in the fundamental IRR's of the
gauge group, this selects only two groups as possible unified models
beyond the electroweak standard model. In particular, they turn out to
be $SU(4)_{PS}\otimes SU(2)_L \otimes SU(2)_R$ and $SU(3)_{C}\otimes
SU(2)_L \otimes SU(2)_R \otimes U(1)_{B-L}$. Remarkably, these groups
appear as the left-right symmetric intermediate stages of $SO(10)$,
although $SO(10)$ itself cannot be realized as a CL model. However,
also these models have troubles since the Dirac operator causing the
correct spontaneous symmetry breaking to the standard model, must
contain Majorana mass terms which spoil gauge invariance of the
theory. 

On one side, it is an appealing feature of this theoretical framework,
the Connes-Lott approach, to be able to select among the possible
gauge groups. On the other hand the fact that only the Standard Model
has survived our analysis is at variance with the currently accepted
idea that new physics beyond the standard model is required at energy
scales lower than Planck mass. Thus this analysis seems to suggest
that a modification is needed either in the basic ingredients of the
model, or in the Hilbert space, which could contain some extra
particles, living at higher energy scales. 

\ \\
{\bf Acknowledgments}\\
We would like to thank F. Buccella, J.M.~Gracia-Bond\'{\i}a and G.~Landi 
and J.C. V\'arilly for useful discussions. 

\def\up#1{\leavevmode \raise.16ex\hbox{#1}}
\newcommand{\npb}[3]{{\sl Nucl. Phys. }{\bf B#1} \up(19#2\up) #3}
\newcommand{\plb}[3]{{\sl Phys. Lett. }{\bf #1B} \up(19#2\up) #3}
\newcommand{\revmp}[3]{{\sl Rev. Mod. Phys. }{\bf #1} \up(19#2\up) #3}
\newcommand{\sovj}[3]{{\sl Sov. J. Nucl. Phys. }{\bf #1} \up(19#2\up) #3}
\newcommand{\jetp}[3]{{\sl Sov. Phys. JETP }{\bf #1} \up(19#2\up) #3}
\newcommand{\rmp}[3]{{\sl Rev. Mod. Phys. }{\bf #1} \up(19#2\up) #3}
\newcommand{\prd}[3]{{\sl Phys. Rev. }{\bf D#1} \up(19#2\up) #3}
\newcommand{\ijmpa}[3]{{\sl Int. J. Mod. Phys. }{\bf A#1} \up(19#2\up) #3}
\newcommand{\prl}[3]{{\sl Phys. Rev. Lett. }{\bf #1} \up(19#2\up) #3}
\newcommand{\physrep}[3]{{\sl Phys. Rep. }{\bf #1} \up(19#2\up) #3}
\newcommand{\jgp}[3]{{\sl J. Geom. Phys.}{\bf #1} \up(19#2\up) #3}
\newcommand{\journal}[4]{{\sl #1 }{\bf #2} \up(19#3\up) #4}


\begin{thebibliography}{99} 
\bibitem{ConnesLott} A.~Connes and J.~Lott, {\sl Nucl.\ Phys.} (Proc.\ Suppl.) 
{\bf B18} (1990) 29.
\bibitem{Connes} A.~Connes, \journal{J.\ Math.\ Phys.}{36}{95}{6194}.
\bibitem{BOOK} A.~Connes, {\it Noncommutative Geometry}, Academic Press, 1994. 
\bibitem{Varilly} J.C.~V\'arilly and J.M.~Gracia-Bond\'{\i}a, \jgp{12}{93}{223}.
\bibitem{Madrid} E.~Alvarez, J.M.~Gracia-Bond\'{\i}a and C.P.~Mart\'{\i}n, 
\plb{306}{93}{55};\plb{329}{94}{259}.
\bibitem{Kastler} D.~Kastler \journal{Rev.\ Math.\ Phys.}{5}{93}{477}.
\bibitem{SchuckZyl} T.~Schucker and J.-M.~Zylinski, hep-th/9312186.
\bibitem{KastlerSchuck} D.~Kastler and T.~Schucker, hep-th/9412185;
hep-th/9501077.
\bibitem{IochSchuck} B.~Iochum and T.~Schucker, hep-th/9501142.
\bibitem{KastIochSchuck} B.~Iochum, D.~Kastler and T.~Schucker, 
\journal{J.\ Math.\ Phys.}{36}{95}{6232}; hep-th/9507150.
\bibitem{cordelia} C.P.~Mart\'{\i}n, J.M.~Gracia-Bond\'{\i}a and 
J.C.~V\'arilly, {\em The Standard Model as a Noncommutative Geometry}, 
to appear.
\bibitem{Becca} B.~Asquith, hep-th/9509163.
\bibitem{Napolinfla} F.~Lizzi, G.~Mangano, G.~Miele and G.~Sparano 
gr-qc/9503040, to appear in {\sl Int.\ J.\ Mod.\ Phys.} {\bf A}.
\bibitem{Zurich1} A.H.~Chamseddine, G.~Felder and J.~Fr\"ohlich, \plb{296}
{93}{109}; \npb{395}{93}{672}. 
\bibitem{Zurich2} A.H.~Chamseddine and J.~Fr\"ohlich, \prd{50}{94}{2893};
\plb{314}{93}{308}.
\bibitem{Pepenu} J.M.~Gracia-Bond\'{\i}a, \plb{351}{95}{510}.
\bibitem{tomita} M.~Tomita, Proc.\ of the {\it Vth functional analysis symposium
of the Math.\ Soc.\ of Japan}, Sendai, 1967.
\bibitem{take} M.\ Takesaki, {\it Tomita's theory of modular Hilbert algebras 
and its applications}, Lecture Notes in Math.\ No.128, Springer-Verlag, 1970.
\bibitem{unif} G.G.~Ross {\it Gran Unified Theories}, Benjamin/Cummings, 
1984; R.~Slansky \physrep{79}{81}{1}.
\bibitem{Pati} J.C.~Pati and A.~Salam, \prd{10}{74}{275}
\bibitem{Madridanom} E.~Alvarez, J.M.~Gracia-Bond\'{\i}a and C.P.~Mart\'{\i}n,
\plb{364}{95}{33}.
\end{thebibliography}
\end{document}